\documentclass[aps,showpacs,preprintnumbers,amsmath, amssymb]{revtex4}

\usepackage{float}
\usepackage{graphics,epsfig}
\usepackage{graphicx}
\usepackage{dcolumn}
\usepackage{bm}

\begin{document}
\baselineskip=0.8 cm

\title{{\bf Effects of quintessence on holographic p-wave superconductors}}
\author{Songbai Chen}
\email{csb3752@hunnu.edu.cn} \affiliation{ Institute of Physics and
Department of Physics, Hunan Normal University,  Changsha, Hunan
410081, P. R. China \\ Key Laboratory of Low Dimensional Quantum
Structures \\ and Quantum Control of Ministry of Education, Hunan
Normal University, Changsha, Hunan 410081, P. R. China}
\author{Qiyuan Pan}
\email{panqiyuan@126.com} \affiliation{ Institute of Physics and
Department of Physics, Hunan Normal University,  Changsha, Hunan
410081, P. R. China \\ Key Laboratory of Low Dimensional Quantum
Structures \\ and Quantum Control of Ministry of Education, Hunan
Normal University, Changsha, Hunan 410081, P. R. China}

\author{Jiliang Jing}
\email{jljing@hunnu.edu.cn}
 \affiliation{ Institute of Physics and
Department of Physics, Hunan Normal University,  Changsha, Hunan
410081, P. R. China \\ Key Laboratory of Low Dimensional Quantum
Structures \\ and Quantum Control of Ministry of Education, Hunan
Normal University, Changsha, Hunan 410081, P. R. China}

\vspace*{0.2cm}
\begin{abstract}
\baselineskip=0.6 cm
\begin{center}
{\bf Abstract}
\end{center}
We construct a holographic p-wave superconductor model in the background of quintessence AdS black hole with an SU(2) Yang-Mills gauge field and then probe the effects of quintessence on the holographic p-wave superconductor.  We investigate the relation between the critical temperature and the state parameter of quintessence, and present the numerical results for electric conductivity. It is shown that the condensation of the vector
field becomes harder as the absolute value of the state parameter increases. Unlike the scalar condensate in the s-wave model, the condensation of the vector
field in p-wave model can occur in the total value range of the state parameter $w_q$ of quintessence. These results could help us know more about holographic superconductor and dark energy.

\end{abstract}

\pacs{11.25.Tq,  04.70.Bw, 74.20.-z} \maketitle
\newpage
\section{Introduction}

It is undoubted that our Universe is undergoing an accelerated
expansion. Within the framework of Einstein¡¯s
general relativity, this late time cosmic
acceleration  could be driven by a yet unknown dark energy, which is an exotic energy component with negative pressure and  currently accounts for $70\%$ of the total mass-energy of the universe. This means that dark energy dominates the evolution of the present Universe and plays an important role in the modern physical cosmology and astronomy. There are several candidates for the dark energy: the vacuum
energy (the cosmological constant) \cite{1a} and dynamical components (such as quintessence \cite{2a},
k-essence \cite{3a} and phantom models \cite{4a}). Models of dark energy differ with respect to the size of the parameter $w$ namely, the relation between the pressure and energy density of the dark energy. For quintessence, $w$ stays in the range $-1\leq w<0$. Obviously, the cosmological constant can be treated as a special type of quintessence. 

Black hole is another important object with very strong gravity. It is of great interest to study the interaction between dark energy and black hole. E. Babichev and his co-workers \cite{EBab} found that all black holes lose their masses in the process of absorbing the phantom dark energy, which implies that the cosmic censorship conjecture is threaten seriously because the charge of a Reissner-Nordstr\"{o}m-like black hole absorbing the phantom energy will be larger than its mass.  We studied the wave dynamics of the phantom scalar perturbation
in the Schwarzschild black hole spacetime and that in the late-time evolution the phantom scalar perturbation grows with an exponential rate rather than decays as the usual scalar perturbations \cite{Sb3,bw2}.  Kiselev \cite{Kiselev} obtained a static solution of Einstein's field equations which describes a black hole surrounded by static and spherically
symmetric quintessence and find that the structure of the spacetime
depends on the state parameter $w_q$ of quintessence. Subsequently,
we \cite{Sb1} extend it to the high dimensional case and find that the temperature and entropy of black hole are functions of the state parameter $w_q$, but the first law is universal. The quasinormal modes and Hawking radiation of these black holes surrounded by quintessence have been studied entirely in \cite{Sb1,Sb2,gyx},  which shows that the presence of quintessence changes the quasinormal frequencies and modifies the standard results of Hawking radiation of the black hole.
These results could help us to get a deeper understand about dark energy and black hole physics.

Recently, holographic superconductor has been attracted much attention since it could provide some insights into the
pairing mechanism in the high $T_c$ superconductors. The AdS/CFT correspondence \cite{ads1} tells us that a gravity theory in a weakly curved $(d+1)$-dimensional AdS spacetimes can be related to the strong coupled $d$-dimensional field theory living at the AdS boundary. With this holographical duality, one can find that black holes with charged scalar hair in AdS spacetimes can provide a holographically dual description of superconductivity. There have been lots of work studying various holographic superconductors \cite{Hs0,Hs01,Hs02,Hsf0, Hsf01,a0,a01,a1,a2,a3,a401,a40,a42,a5,a6,a7,a8,a8d1,GT,a81,a9,a10,a100,
a11,a12,a13,a14,a15,a151,a17,GSJ,gx,bw1,Rgc,Jing1,wuj1} in which dependence of holographic superconductors on the background spacetime, the electrodynamics, condensate fields and back reaction have been discussed. It is shown that the ratio of gap frequency over critical temperature $\omega_g/T_c$ is found to be larger than that in the weakly correlated BCS theory, which implies that the coupling in the holographic superconductor is more strong.

It is of great interest to extend of the study of dark energy to holographic superconductor since it could help us to probe the relationship among the dark energy, black hole and condensed matter physics. Recently, we \cite{Sb5} study the effect of quintessence of s-wave holographic superconductor and find that the larger absolute value
of $w_q$ leads to the lower critical temperature $T_c$ and the
higher ratio between the gap frequency in conductivity to the
critical temperature for the scalar condensates. Moreover, for the scalar condensate there exists an additional constraint
condition originating from the quintessence
$(d-1)w_q+\lambda_{\pm}>0$ for the operators $\mathcal{O}_{\pm}$,
respectively. These results imply that the presence of dark energy will bring richer physics for the holographic superconductor. The p-wave superconductor model \cite{a4} is another important model of holographic superconductor in which there is a special direction
which breaks the rotational symmetry and the conductivity is anisotropic. The properties of p-wave holographic superconductor in the Gauss-Bonnet gravity and in Quasi-topological gravity have been studied in \cite{Rcg1,KLY}. 
In this paper, we are interested in the effect of the quintessence on the p-wave superconductors and to see what effect of the state parameter $w_q$ on the critical temperature, the condensation formation and conductivity.

This paper is organized as follows. In Sec. II, we review briefly the
metric describing a $4$-dimensional planar quintessence AdS black
hole and construct a p-wave holographic superconductor model. In Sec. III, we apply the
Sturm-Liouville analytical \cite{GSJ} and numerical methods to
explore the relations between the critical temperature and the state
parameter $w_q$ of quintessence. In Sec. IV, we probe the effect of
the state parameter on the electrical conductivity of the charged
condensates. Finally in the last section we will include our
conclusions.

\section{Holographic p-wave dual in the quintessence AdS black hole spacetime}

In this section, we review briefly the metric describing a
$4$-dimensional planar quintessence AdS black hole and build the
holographic p-wave dual of the quintessence AdS black hole by
introducing Yang-Mills field. The metric of the quintessence AdS black hole is given by
\begin{eqnarray}
ds^2=-f(r)dt^2+\frac{1}{f(r)}dr^2+r^2(dx^2+dy^2),\label{m1}
\end{eqnarray}
with
\begin{eqnarray}
f(r)=\frac{r^2}{L^2}-\frac{M}{r^{3w_q+1}}.
\end{eqnarray}
$L$ is the radius of AdS and $w_q$ is the state parameter of quintessence. $M$ is the mass parameter of black hole related to the  energy density $\rho_q$ of quintessence by
\begin{eqnarray}
\rho_q=\frac{-3Mw_q}{r^{3(w_q+1)}}.
\end{eqnarray}
Here the energy density $\rho_q$ is positive since the state parameter $w_q<0$ for quintessence. Obviously, the spacetime of black hole depends on the state parametric of quintessence $w_q$. The
Hawking temperature of the quintessence AdS black hole is
given by
\begin{eqnarray}
T_H=\frac{3(w_q+1)r_H}{4\pi L^2},
\end{eqnarray}
which is a function of the state parameter $w_q$ of quintessence. As the state parametric of
quintessence $w_q=0$,  the metric (\ref{m1}) can be reduced to the
usual planar Schwarzschild AdS black hole. 

Let us now consider Einstein-Yang-Mills theory with a Lagrangian
\begin{eqnarray}
\mathcal{L}=-\frac{1}{4}F^{a}_{\mu\nu}F^{a\mu\nu},
\end{eqnarray}
where
$F^{a}_{\mu\nu}=\partial_{\mu}A^a_{\nu}-\partial_{\nu}A^a_{\mu}+\epsilon^{abc}A^a_{\mu}A^b_{\nu}$
is the Yang-Mills field strength, $\epsilon^{abc}$ is the totally
antisymmetric tensor with $\epsilon^{123}=+1$. Following Ref. \cite{a4}, we here adopt the ansatz
\begin{eqnarray}
A_{\mu}=\phi(r)\tau^3dt+\psi(r)\tau^1dx,\label{ansatz}
\end{eqnarray}
for Yang-Mills gauge field. Here $\tau^i$ is are the three $SU(2)$
generators with commutation relation $[\tau^i,
\tau^j]=\epsilon^{ijk}\tau^k$.  In this ansatz one can regard the
$U(1)$ subalgebra generated by $\tau^3$ as the gauge group of
electromagnetism, and the gauge boson with nonzero component
$\psi(r)$ along $x$ direction is charged under $A^3_t=\phi(r)$. From
AdS/CFT correspondence,  $\phi(r)$ is dual to the chemical potential
in the boundary field theory and $\psi(r)$ is dual to the $x$
component of some charged vector operator $J$. The condensation of
$\psi(r)$ breaks the rotational $U(1)$ symmetry and leads to a phase
transition, which can be interpreted as a p-wave superconducting
phase transition on the boundary.

The Yang-Mills equations with the above ansatz (\ref{ansatz}) are
\begin{eqnarray}
&&\psi^{''}+\frac{f'}{f}\psi' +\frac{\phi^2\psi}{f^2}=0,\label{e1}\\
&&\phi^{''}+\frac{2}{r}\phi' -\frac{\psi^2}{r^2f}\phi=0,\label{e2}
\end{eqnarray}
Here we set $L^2=1$ and a prime denotes the derivative with respect
to $r$. In order to solve the equations (\ref{e1}) and (\ref{e2}),
we need the boundary conditions for the fields $\psi(r)$ and
$\phi(r)$. The regularity condition at the horizon gives the
boundary conditions
\begin{eqnarray}
&&\phi=\phi^{(1)}_H(1-\frac{r_h}{r})+... ,\nonumber\\
&&\psi=\psi^{(0)}_H+\psi^{(2)}_H(1-\frac{r_h}{r})^2+... .
\end{eqnarray}
At the spatial infinite $r\rightarrow\infty$, the filed $\psi$ and
$\phi$ behave like
\begin{eqnarray}
&&\psi(r)=\psi^{(0)}+\frac{\psi^{(1)}}{r},\label{b1}\\
&&\phi(r)=\mu+\frac{\rho}{r},\label{b2}
\end{eqnarray}
The constants $\mu$ and $\rho$ can be interpreted as the chemical
potential and charge density in the dual field theory, respectively.
For the holographic p-wave superconductor, the normal charge density
$\rho_n=\phi^{(1)}_H$, and the total charge density is
$\rho_t=2\rho$. So the superconducting charge density is given by
$\rho_s=\rho_t-\rho_n$. The coefficients $\psi^{0}$ and $\psi^{1}$
are dual to the source and expectation value of the boundary
operator $J^{1}_x$ respectively. For normalizable modes, the
expectation value of vacuum can be obtained by setting the source
$\psi^{0}$ to zero.

\section{Properties of holographic p-wave superconductors in the quintessence AdS black hole spacetime}

In this section we will use both the Sturm-Liouville analytical
\cite{GSJ} and numerical methods to study properties of holographic
p-wave superconductors in the quintessence AdS black hole spacetime
and probe the effect of the state parameter of quintessence $w_q$ on
the holographic p-wave superconductors.

Introducing a new coordinate $z=r_H/r$, the equations of motion
(\ref{e1}) and (\ref{e2}) for two fields $\psi$ and $\phi$ can be
rewritten as
\begin{eqnarray}
&&\psi^{''}+\bigg(\frac{f'}{f}+\frac{2}{z}\bigg)\psi'
+\frac{\phi^2r^2_H}{z^4f^2}\psi=0,\label{e3}\\
&&\phi^{''}-\frac{\psi^2}{z^2f}\phi=0,\label{e4}
\end{eqnarray}
where a prime denotes the derivative with respect to $z$. When
$T\rightarrow T_c$, one can find the condensation tends to zero,
i.e., $\psi\rightarrow0$. From Eq.(\ref{e4}), one can find that near
the critical temperature the field $\phi$ can be expressed as
\begin{eqnarray}
\phi=\xi r_H(1-z),
\end{eqnarray}
where $\xi=\rho/r^2_H$. Near the boundary $z=0$, one can introduce a
trial function $F(z)$ which is related to $\psi$ by
\begin{eqnarray}
\psi\sim\frac{\psi^{(1)}}{r}\sim\langle J^{1}_x
\rangle\frac{z}{r_H}F(z).
\end{eqnarray}
The trial function $F(z)$ should satisfy the boundary condition
$F(0)=1$ and $F'(0)=0$. Then we can get the equation for $F(z)$
\begin{eqnarray}
F''(z)&+&\bigg[\frac{2-(3w_q+5)z^{3(w_q+1)}}{z(1-z^{3(w_q+1)})}\bigg]F'(z)
-\frac{3(w_q+1)z^{3w_q+1}}{(1-z^{3(w_q+1)})}F(z)+
\frac{\xi^2(1-z)^2}{(1-z^{3(w_q+1)})^2}F(z)=0.\label{Fz1}
\end{eqnarray}
Multiplying the above equation with the following function
\begin{eqnarray}
T(z)=z^{2}(z^{3(w_q+1)}-1),
\end{eqnarray}
we can rewritten Eq. (\ref{Fz1}) as
\begin{eqnarray}
[T(z)F'(z)]'-Q(z)F(z)+\xi^2P(z)F(z)=0,\label{Fz2}
\end{eqnarray}
with
\begin{eqnarray}
Q(z)=T(z)\frac{3(w_q+1)z^{3w_q+1}}{(1-z^{3(w_q+1)})},\;\;\;\;\;
P(z)=T(z)\frac{(1-z)^2}{(1-z^{3(w_q+1)})^2}.
\end{eqnarray}
Making use of the Sturm-Liouville method, one can obtain the minimum
eigenvalue of $\xi^2$ in Eq. (\ref{Fz2}) can be calculated by
\begin{eqnarray}
\xi^2=\frac{\int^{1}_{0}[T(z)F'(z)^2+Q(z)F(z)^2]dz}{\int^{1}_{0}P(z)F(z)^2dz}.
\end{eqnarray}
The trial function $F(z)$ satisfied its boundary condition can be
taken as $F(z)=1-az^2$, so $\xi^2$ can be explicitly written as
\begin{eqnarray}
\xi^2=\frac{s(w_q,a)}{t(w_q,a)}.
\end{eqnarray}
For different values of $w_q$, $d$ and $m$, we can obtain the
minimum value of $\xi^2$ with appropriate value of $a$.

Combing $T=\frac{3(w_q+1)r_H}{4\pi}$ and $\xi=\frac{\rho}{r^2_H}$,
we can obtain the form of the critical temperature $T_c$
\begin{eqnarray}
T_c=\gamma\sqrt{\rho},
\end{eqnarray}
with the coefficient $\gamma=\frac{w_q+1}{4\pi\sqrt{\xi_{min}}}$.
Thus, we can probe the effects of the state parameter $w_q$ on the
critical temperature $T_c$ through estimating $\xi_{min}$ by using
the Sturm-Liouville method.

In Table (I), we present the analytical values and numerical
values of critical temperature $T_c$ for the holographic p-wave superconductor with different values of the state parameter $w_q$,
which exhibits that the analytical values of $T_c$ from the
Sturm-Liouville method agree entirely with the exact numerical
results.
\begin{table}[ht]
\begin{tabular}[b]{|c|c|c|c|c|c|}
\hline\hline&&&&& \;\;\;\;\;\; \\
 $w_q$&\;$0$&\;\;$-0.2$&\;\;$-0.4$&\;\;$-0.6$
 &\;\;$-0.8$\\
 \hline&&&&&\\
 $T_c$ (Analytical)&\;$0.1239$&\;\;$0.1019$&\;\;$0.0802$&\;\;$0.0585$
 &\;\;$0.0363$\\
 \hline&&&&&\\
 $T_c$(Numerical)&\;$0.1250$&\;\;$0.1024$&\;\;$0.0803$&\;\;$0.0585$
 &\;\;$0.0366$\\
 \hline\hline
\end{tabular}
\caption{Variety of the critical temperature $T_c$ for the holographic p-wave superconductor with different values of the state parameter $w_q$.}
\end{table}
\begin{figure}[ht]
\begin{center}
\includegraphics[width=7cm]{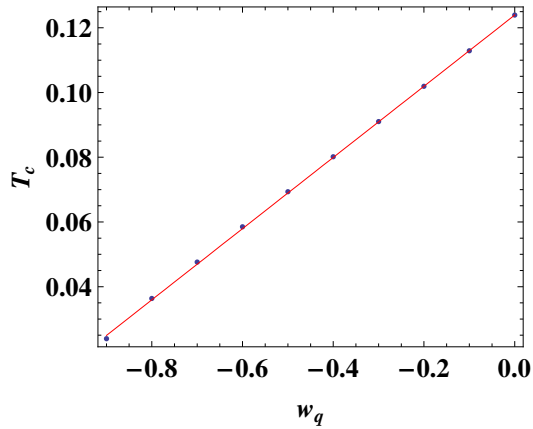}
\caption{Dependence of the critical temperature $T_c$ on the state parameter $w_q$. The solid line denotes the function $T_c=aw_q+b$. We here set the numerical constants $a=0.110$ and $b=0.124$.}
\end{center}
\end{figure}
\begin{figure}[ht]
\begin{center}
\includegraphics[width=7cm]{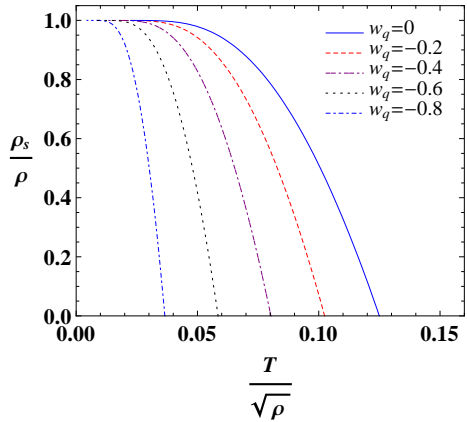}
\caption{The ratio of the superconducting charge density to the total charge density as a function of the temperature with different values of the state parameter $w_q$.}
\end{center}
\end{figure}
\begin{figure}[ht]
\begin{center}
\includegraphics[width=7cm]{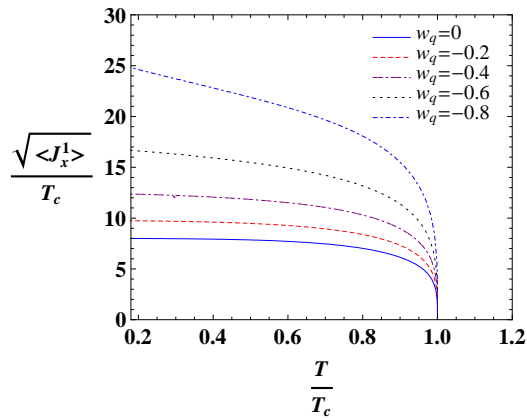}
\caption{The condensate as a function of the temperature with different values of the state parameter $w_q$.}
\end{center}
\end{figure}
Moreover, we find that the dependence of the critical temperature $T_c$ on the state parameter $w_q$ can be fitted best by
\begin{eqnarray}
 T_c=aw_q+b,
\end{eqnarray}
with two numerical constants $a$ and $b$ , which is shown in Fig. (1). It means that $T_c$ decreases linearly with the absolute value of $w_q$.
In Fig. (2), we plot the ratio $\rho_s/\rho_t$ v.s. $T/\sqrt{\rho}$ by changing the state parameter $w_q$. The points where the curves intersect with the horizontal axis represent the critical temperatures for different $w_q$ when the superconducting phase occurs. It is clear that the critical temperature decreases with the absolute value of $w_q$. This means that the occurrence of condensation should
become more difficultly in the case with a larger absolute value of $w_q$. It could be explained by the fact that the quintessence with a larger absolute value of $w_q$ possesses the stronger negative pressure which yields the condensation become harder. In Fig. (3), we illustrate the condensation of $J^1_x$
as the function of the temperature with different values of the state parameter.
It is shown that the order parameter has the behavior $
\langle J^1_x\rangle\sim(1-T/T_c)^{1/2}$ near the critical temperature,
and the value of $\sqrt{\langle J^1_x\rangle}/T_c$
increases with the state parameter $w_q$. These effects of the state parameter $w_q$ on holographic p-wave superconductor are similar to the effects of $w_q$ on holographic s-wave superconductor in the AdS quintessence black hole. However, we also find that in the p-wave case the condensate always form in the total value range of the state parameter $w_q$ of quintessence. It is different from that in the s-wave case in which there does exist an additional constraint condition
originating from the quintessence for the scalar condensate.
In the next section, we will calculate the frequency dependent conductivity to see the influence of the quintessence more clearly.

\section{The electrical conductivity}

Let us now investigate the influence of the state
parameter $w_q$ of quintessence on the electrical conductivity.
In order to calculate the electric conductivity of the system, we must introduce an electromagnetic perturbation into the system. For the Yang-Mills case, the perturbation of the electric field can be taken as \cite{a4}
\begin{eqnarray}
\delta A=e^{-i\omega t}[(A^{1}_t(r)\tau^1+A^{2}_t(r)\tau^2)dt+A^3_x(r)\tau^3dx
+A^3_y(r)\tau^3dy]. \label{Ewr}
\end{eqnarray}
Substituting the perturbation (\ref{Ewr}) into the linearized Yang-Mills
equations, one can find the equation of motion for $A^3_y(r)$ is decoupled from other components of the Yang-
Mills field, but for $A^3_x(r)$ it mixes with other two components $A^1_t(r)$ and $A^2_t(r)$. This implies that the properties of the electric conductivity component $\sigma_{xx}$ are different from those for the component $\sigma_{yy}$. In the following subsections, we separately calculate the components $\sigma_{yy}$ and $\sigma_{xx}$ to see the difference between them, and study the effect of the state
parameter $w_q$ of quintessence on the electric conductivities.

\subsection{$\sigma_{yy}$}

The equation of motion for $A^3_y$ can be expressed as
\begin{eqnarray}
A^{3''}_y+\frac{f'}{f}A^{3'}_y
+\bigg[\frac{\omega^2}{f^2}-\frac{2\psi^2}{f}\bigg]A^3_y=0.\label{de}
\end{eqnarray}
Obviously, it is very similar to corresponding equation in the holographic s-wave superconductors. Thus, $\sigma_{yy}$ can be calculated by the same method applied in the s-wave case. We choose the ingoing
wave condition for $A^3_y$ near the horizon, and then we have
\begin{eqnarray}
A^3_y= (1-\frac{r_H}{r})^{-\frac{i\omega }{3(w_q+1)r_H}}[1+a^{3(1)}_y(1-\frac{r_H}{r})+a^{3(2)}_y(1-\frac{r_H}{r})^2+...].\label{bd1}
\end{eqnarray}
\begin{figure}[ht]
\begin{center}
\includegraphics[width=6.8cm]{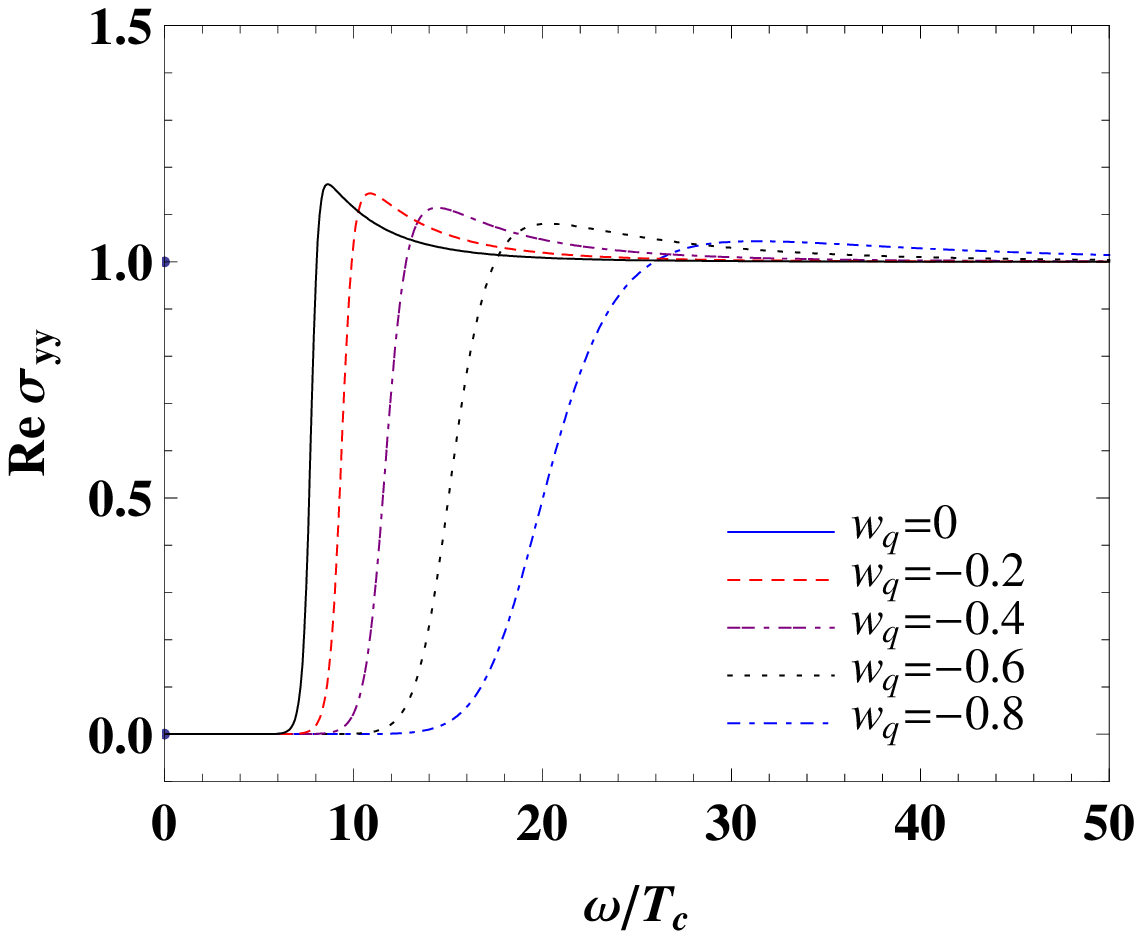}\;\;\;\includegraphics[width=6.8cm]{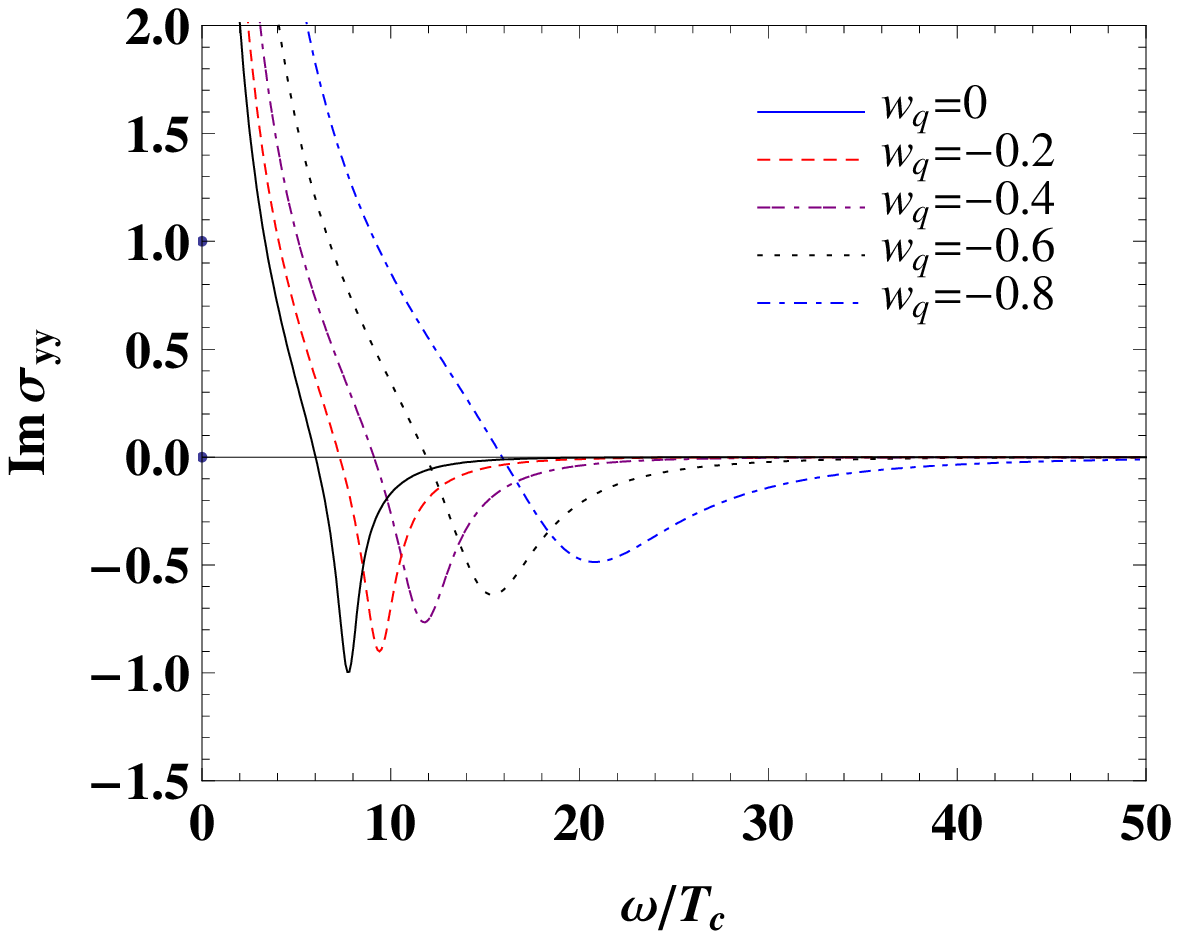}\\
\caption{The conductivity $\sigma_{yy}$ for the p-wave superconductors with
different values of $w_q$ in the quintessence AdS
black hole spacetime. The left figure is for the real part of $\sigma_{yy}$  while the right one is for the imaginary part of $\sigma_{yy}$. Each plot is at low temperatures about $T/T_c\approx 0.19$.}
\end{center}
\end{figure}
At the asymptotic AdS region, the general behavior of $A^3_y$ can
be expressed as
\begin{eqnarray}
A^3_y=A^{(0)}_y+\frac{A^{(1)}_y}{r},\label{bd2}
\end{eqnarray}
From the AdS/CFT dictionary, the conductivity can be expressed through the retarded
Green function as follows \cite{Hs0}
\begin{eqnarray}
\sigma_{yy}=\frac{1}{i\omega}G^{R}(\omega)|_{k=0}=-i\frac{A^{3(1)}_y}{\omega A^{3(0)}_y}.
\end{eqnarray}
With the boundary conditions (\ref{bd1}) and (\ref{bd2}), we numerically solve equation (\ref{de}) and obtain $\sigma_{yy}$. In the Fig. (3), we plotted the dependent relation between $\sigma_{yy}$ and $\omega$ for different values of $w_q$.
It is clearly that with the decrease of the state parameter $w_q$ the real part of the conductivity $\sigma_{yy}$ decreases at the lower frequency and increases at the higher frequency. The variety of the imaginary part of $\sigma_{yy}$ with $w_q$ is  converse to the variety of the real part of $\sigma_{yy}$ with $w_q$. Moreover, one can find that there exists a minimum for the imaginary part of $\sigma_{yy}$ at $\omega=\omega_g$.  The gap frequency $\omega_g$ describes excitation of quasi-particles in pairs and can
be interpreted as the energy to break a pair of fermions. From the right panel of Fig.(4), we find that
the ratio of the gap frequency $\omega_g$ over the critical temperature $T_c$ increases when the absolute value of the state parameter becomes more large, which means that the
occurrence of p-wave holographic superconductor needs the stronger coupling
in the quintessence AdS black hole spacetime. This behavior is much similar to that of $\sigma$ in the s-wave case.

\subsection{$\sigma_{xx}$}

We are now in position to study the component $\sigma_{xx}$, which can be obtained by solving the equation of motion for $A^3_x$ in the bulk. In contrast to $\sigma_{yy}$, the calculation of $\sigma_{xx}$ is more complicated since in the linearized Yang-Mills equations the component $A^3_x$ couples with other components
$A^1_t$ and $A^2_t$. The equations of motion for the three components
are
\begin{eqnarray}
&&A^{3''}_x+\frac{f'}{f}A^{3'}_x
+\frac{\omega^2}{f^2}A^3_x-\frac{i\omega A^2_t+A^1_t\phi}{f^2}\psi=0,\label{dex1}\\
&&A^{1''}_t+\frac{2}{r}A^{1'}_t
+\frac{A^3_x\phi\psi}{r^2f}=0,\label{dex2}\\
&&A^{2''}_t+\frac{2}{r}A^{2'}_t-\frac{A^2_t\psi^2}{r^2f}-\frac{i\omega A^3_x\psi}{r^2f}=0.\label{dex3}
\end{eqnarray}
Moreover, there are two additional constraints
\begin{eqnarray}
i\omega A^{1'}_t+\phi A^{2'}_t- A^{2}_t\phi'&=&0,\\
i\omega A^{2'}_t-\phi A^{1'}_t+A^{1}_t\phi'-\frac{f\psi A^{3'}_x}{r^2}+\frac{f\psi' A^{3}_x}{r^2}&=&0.
\end{eqnarray}
Similarly, these constraints are not independent of the equations of
motion (\ref{dex1}), (\ref{dex2}) and (\ref{dex3}). Again using the ingoing wave condition near the horizon, we can find that
\begin{eqnarray}
A^3_x&=&(1-\frac{r_H}{r})^{-\frac{i\omega }{3(w_q+1)r_H}}[1+a^{3(1)}_x(1-\frac{r_H}{r})+a^{3(2)}_x(1-\frac{r_H}{r})^2+...],
\label{bdx1}\\
A^1_t&=&(1-\frac{r_H}{r})^{-\frac{i\omega }{3(w_q+1)r_H}}[a^{1(2)}_t(1-\frac{r_H}{r})^2+a^{1(3)}_t(1-\frac{r_H}{r})^3+...],
\label{bdx2}\\
A^2_t&=&(1-\frac{r_H}{r})^{-\frac{i\omega }{3(w_q+1)r_H}}[a^{2(1)}_t(1-\frac{r_H}{r})+a^{2(2)}_t(1-\frac{r_H}{r})^2+...].
\label{bdx3}
\end{eqnarray}
Here $a^{a(i)}_{\mu}$ are some constants, which depend on the state parameter $w_q$ of the quintessence and the frequency $\omega$ of the electromagnetic perturbation.
Near the boundary of the AdS bulk ($r\rightarrow \infty$), we have
\begin{eqnarray}
A^3_x&=&A^{3(0)}_x+\frac{A^{3(1)}_x}{r},
\label{bdx21}\\
A^1_t&=&A^{1(0)}_t+\frac{A^{1(1)}_t}{r},
\label{bdx22}\\
A^2_t&=&A^{2(0)}_t+\frac{A^{2(1)}_t}{r}.
\label{bdx23}
\end{eqnarray}
These coefficients in the expansions can be determined numerically. Since these quantities are gauge dependent, we need define a gauge invariant quantity to calculate the conductivity $\sigma_{xx}$. Following Ref.\cite{a4}, one can find the gauge invariant quantity should be
\begin{eqnarray}
\hat{A}^3_x=A^3_x+\psi\frac{i\omega A^2_t+\phi A^1_t}{\phi^2-\omega^2},
\end{eqnarray}
which is a linear combination of $A^3_x$, $A^1_t$ and $A^2_t$.
This gauge invariant quantity near the AdS boundary behaves like
\begin{figure}[ht]
\begin{center}
\includegraphics[width=7.2cm]{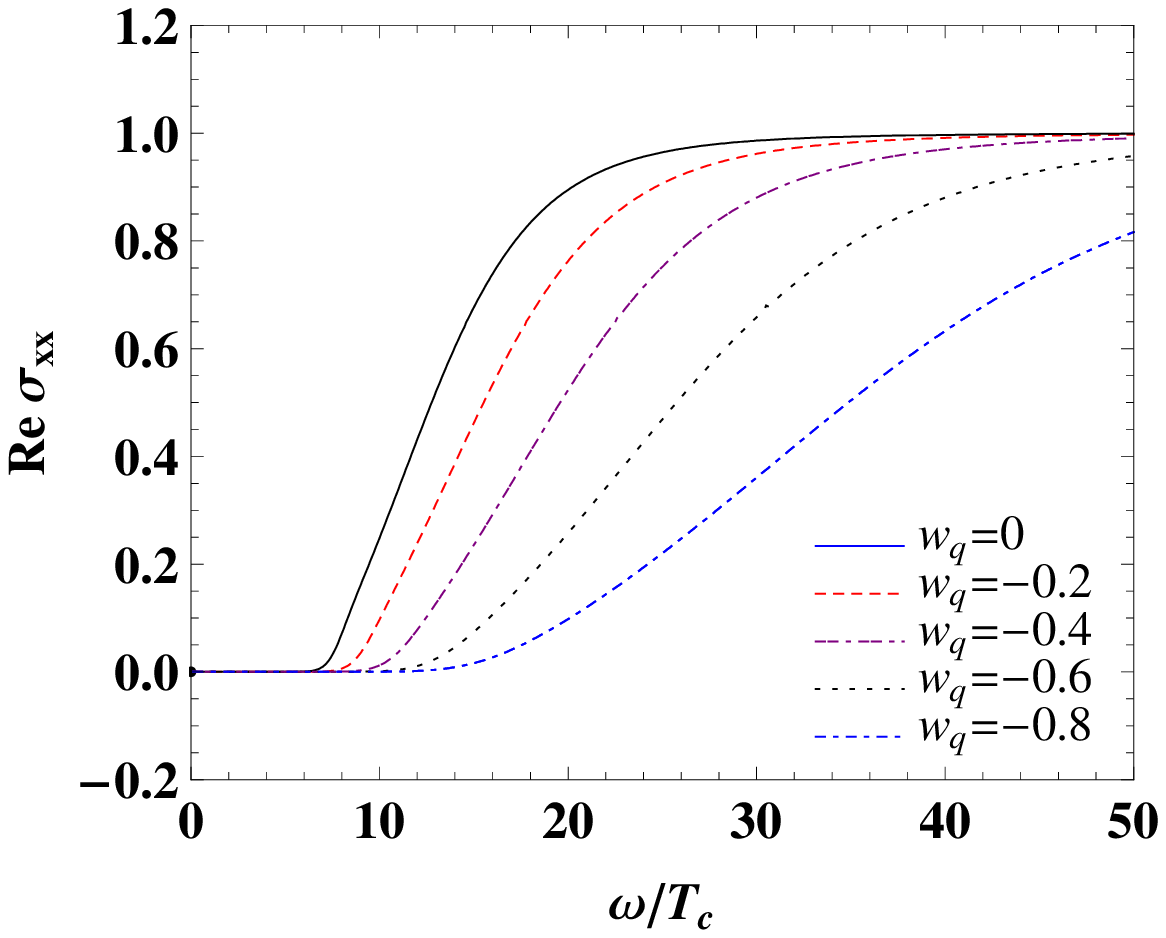}\;\;\;\includegraphics[width=6.7cm]{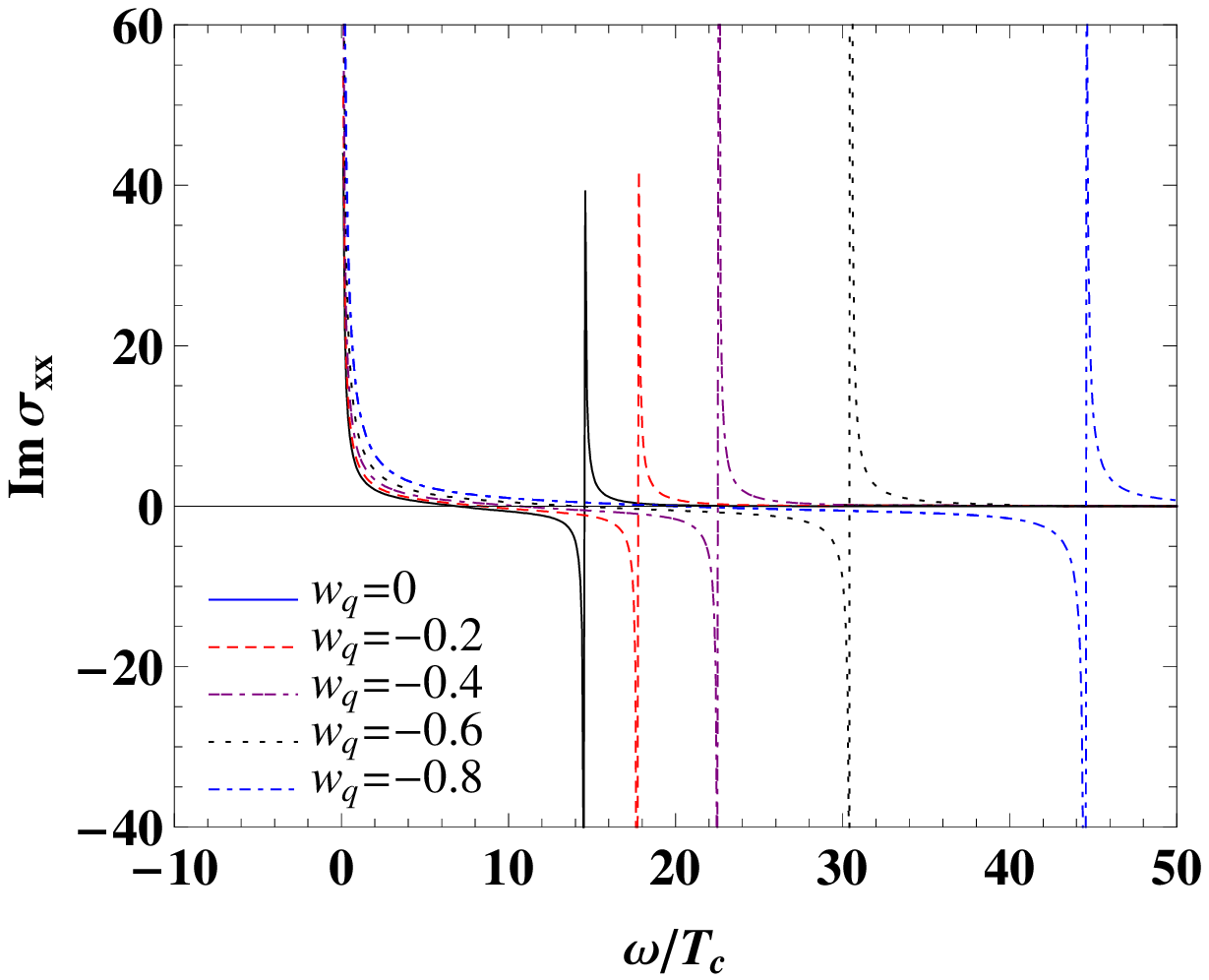}\\
\caption{The conductivity $\sigma_{xx}$ for the p-wave superconductors with
different values of $w_q$ in the quintessence AdS
black hole spacetime. The left figure is for the real part of $\sigma_{yy}$  while the right one is for the imaginary part of $\sigma_{xx}$. Each plot is at low temperatures about $T/T_c\approx 0.19$.}
\end{center}
\end{figure}
\begin{eqnarray}
\hat{A}^3_x=A^{3(0)}_x+\frac{\hat{A}^{3(1)}_x}{r}+...,
\end{eqnarray}
with
\begin{eqnarray}
\hat{A}^{3(1)}_x=A^{3(1)}_x+\psi^{(1)}\frac{i\omega A^{2(0)}_t+\mu A^{1(0)}_t}{\mu^2-\omega^2}.
\end{eqnarray}
In doing so, one can obtain the conductivity $\sigma(\omega)_{xx}$ by \cite{a4}
\begin{eqnarray}
\sigma_{xx}=-i\frac{\hat{A}^{3(1)}_x}{\omega A^{3(0)}_x}=-\frac{i}{\omega A^{3(0)}_x}\bigg[A^{3(1)}_x+\psi^{(1)}\frac{i\omega A^{2(0)}_t+\mu A^{1(0)}_t}{\mu^2-\omega^2}\bigg].
\end{eqnarray}
In Fig.(5), we present the numerical results of the electric conductivity  $\sigma_{xx}$. It is shown that the point of phase transition also changes as we choose different value of the state parameter $w_q$ of quintessence. Moreover, one can find that the real part of $\sigma_{xx}$ decreases with the absolute value of the state parameter $w_q$ in the total range of the frequency spectrum $\omega$. It means that the dependence of $\sigma_{xx}$ on the state parameter $w_q$ is different from the dependence of $\sigma_{yy}$ on the state parameter $w_q$. Actually, this anisotropic behavior of conductivity is just the feature of p-wave
superconductors. For the imaginary part of $\sigma_{xx}$, we can find that
there also exists a pole, which moves along right with the increase of the absolute value of the state parameter $w_q$.

\section{Summary}

In this paper we have constructed a p-wave holographic superconductor model in
a $4$-dimensional planar quintessence AdS black hole spacetime in the probe limit and probe the effect of the state parameter $w_q$ of quintessence on the state parameter
$w_q$ on the critical temperature $T_c$ and the condensation
formation and conductivity. Our results show that the critical temperature $T_c$ decreases linearly with the absolute value of the state parameter $w_q$ of quintessence, which means that the presence of quintessence makes the form of the p-wave superconducting condensation more hard. This behavior is very similar to the case of the s-wave model. Moreover, we also find that the holographic p-wave superconductor can occur in the total value range of the state parameter $w_q$ of quintessence and there does not exist such an additional constraint condition originating from the quintessence (i.e., $3w_q+\lambda_{\pm}>0$) as in the s-wave model.

We also investigate anisotropic conductivities $\sigma_{xx}$ and $\sigma_{yy}$ in the p-wave superconductor. It is shown that the effect of $w_q$ on the $\sigma_{yy}$ is similar to that in the s-wave model. The ratio of
gap frequency $\omega_g$ over the critical temperature $T_c$ increases when the absolute value of the state parameter becomes more large, which means that the
occurrence of p-wave holographic superconductor needs the stronger coupling
in the quintessence AdS black hole spacetime. However, the conductivity $\sigma_{xx}$  behaves much different. The real part of $\sigma_{xx}$ decreases with the absolute value of the state parameter $w_q$ in the total range of the frequency spectrum $\omega$. For the imaginary part of $\sigma_{xx}$, one can find that
there also exists a pole, which moves along right with the increase of the absolute value of the state parameter $w_q$.

\begin{acknowledgments}
This work was  partially supported by the NCET under Grant
No.10-0165, the PCSIRT under Grant No. IRT0964 and the construct
program of key disciplines in Hunan Province. J. Jing's work was
partially supported by the National Natural Science Foundation of
China under Grant Nos. 11175065, 10935013; 973 Program Grant No.
2010CB833004.
\end{acknowledgments}

\end{document}